\documentclass[prb,twocolumn,showpacs,preprintnumbers,amsmath,amssymb,superscriptaddress]{revtex4}
\usepackage{graphicx}
\usepackage{color}
\usepackage{bm}
\usepackage{epstopdf}
\usepackage{color}
\pdfoutput=1

\begin{document}

\title{Bulk properties and electronic structure of PuFeAsO}

\author{T. Klimczuk}
\email{tomasz.klimczuk@pg.gda.pl}%
\affiliation{European Commission, JRC, Institute for Transuranium Elements, Postfach 2340, 76125 Karlsruhe, Germany}
\affiliation{Faculty of Applied Physics and Mathematics, Gdansk University of Technology, Narutowicza 11/12, 80-952 Gdansk, Poland}
\author{A. B.  Shick}
\affiliation{European Commission, JRC, Institute for Transuranium Elements, Postfach 2340, 76125 Karlsruhe, Germany}
\affiliation{Institute of Physics, ASCR, Na Slovance 2, 182 21, Prague, Czech Republic}
\author{R. Springell}
\affiliation{Royal Commission for the Exhibition of 1851 Research Fellow, Interface Analysis Centre, University of Bristol, Bristol BS2 8BS, United Kingdom}
\author{H. C. Walker}
\affiliation{Deutsches Elektronen-Synchrotron (Hasylab at DESY), 22607 Hamburg, Germany}
\author{A. H. Hill}
\altaffiliation[Currently at: ]{Johnson Matthey Technology Centre, Sonning Common, UK}
\affiliation{European Synchrotron Radiation Facility, 6 rue Jules Horowitz, BP220, 38043 Grenoble Cedex 9, France}
\author{E. Colineau}
\author{J.-C. Griveau}
\affiliation{European Commission, JRC, Institute for Transuranium Elements, Postfach 2340, 76125 Karlsruhe, Germany}
\author{D. Bou\"{e}xi\`{e}re}
\author{R. Eloirdi}
\author{R. Caciuffo}
\email{roberto.caciuffo@ec.europa.eu}%
\affiliation{European Commission, JRC, Institute for Transuranium Elements, Postfach 2340, 76125 Karlsruhe, Germany}

\date{\today}

\begin{abstract}
Here we present bulk property measurements and electronic structure calculations for PuFeAsO, an actinide analogue of the iron-based rare-earth superconductors \emph{R}FeAsO. Magnetic susceptibility and heat capacity data suggest the occurrence of an antiferromagnetic transition at $T_N$=50~K. No further anomalies have been observed down to 2~K, the minimum temperature that we have been able to achieve. Structural measurements indicate that PuFeAsO, with its more localized $5f$ electrons, bears a stronger resemblance to the \emph{R}FeAsO compounds with larger \emph{R} ions, than NpFeAsO does.

 \end{abstract}


\maketitle

\section{Introduction}
Strongly correlated materials provide one of the most exciting arenas for theoretical and experimental research in condensed matter physics. Compounds containing elements of the actinide series yield many of the greatest challenges in this field, driven principally by their 5\textit{f} electronic states, which display characteristics that are simultaneously itinerant and localized\cite{Zwicknagl,santini09}. Plutonium is at the crossroads of band-like and atomic-like behavior \cite{Hecker2003}, and compounds containing plutonium display a fantastic range of physical phenomena, as a direct consequence of this unique electronic property \cite{Joyce2003, Zhu2007}.

A huge volume of research has resulted from the recent discovery of superconductivity\cite{Kamihara2006} in the \textit{R}FeAsO series of compounds, where \textit{R} represents a rare-earth element, characterized by its atomic-like 4\textit{f} electrons. The replacement of this species by a member of the actinide series presents an exciting opportunity to study the effect of varying electron correlation. In fact, the successful substitution with Np to form the NpFeAsO parent compound has already been reported \cite{Klimczuk2012}, where the physical properties were fundamentally different from those of the \textit{R}FeAsO family. The replacement of Np with Pu is the next logical step, i.e. the replacement of an itinerant 5\textit{f} band with 5\textit{f} electronic states which straddle the transition metal/rare-earth regime. Here, we present the synthesis, crystal structure, bulk physical properties and the first-principles electronic structure calculations of the first Pu-based oxypnictide, PuFeAsO.

\section{Experimental details}

Polycrystalline samples of PuFeAsO were produced by solid state reaction using PuAs, Fe$_3$O$_4$ and elemental Fe as described in ref. \onlinecite{Klimczuk2012}. The room temperature crystallographic structure was determined by X-ray powder diffraction on a Bruker D8 Focus diffractometer put inside a glove box for measurements on transuranium compounds. The diffractometer is equipped with a Cu $K\alpha_{1}$ source and a germanium (111) monochromator. To avoid contamination of the box, the powder sample was embedded in an epoxy glue.

Measurements of the magnetic susceptibility, $\chi(T)$, have been performed in the temperature interval 2-300 K using a Quantum Design MPMS-7T superconducting quantum interference device (SQUID) and a polycrystalline sample hold in a plastic tube. Raw longitudinal magnetization data have been corrected for diamagnetic contributions and the small additional signal due to the empty sample holder.
Specific heat measurements were performed as a function of temperature using a Quantum Design PPMS-9T system via the relaxation method. Because of the radioactivity
and toxicity of plutonium, the sample was wrapped in the heat conducting
STYCAST $\circledR$ 2850 FT resin to avoid
contamination risks. Due to the self-heating effects associated to the $\alpha$ decay of $^{239}$Pu, it was not possible to extend the heat capacity measurements below $\sim$~7~K.The measured data have been corrected for the encapsulation
contribution by applying
a standard procedure. The contribution to the heat capacity curve due to a  PuO$_2$ impurity phase has been subtracted by using the data reported in ref.~\onlinecite{Flotow1976}.

\section{Results and discussion}

\begin{figure}[t]
\centering
\includegraphics[width=0.45\textwidth,bb=60 10 740 530,clip]{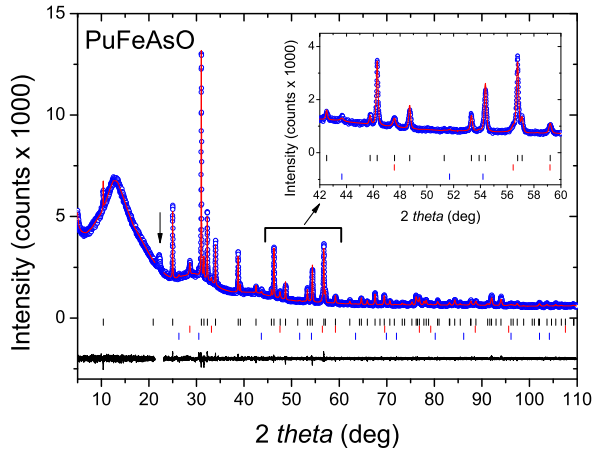}\caption
{\label{xrd}(Color online) Rietveld refinement of room temperature x-ray powder diffraction data for $\mathrm{PuFeAsO}$. Upper part: blue circles - observed data, red solid line - calculated intensities. The lower part shows, on the same scale, the difference between the observed and calculated patterns. The black (upper) tick marks correspond to the Bragg peaks from the $\mathrm{PuFeAsO}$ (\textit{P}4/\textit{nmm}, s.g. 129) structure, with the red (middle) and blue (lower) sets referring to the PuAs and PuO$_2$ impurities of $3\%$ and $6\%$ respectively. The vertical arrow indicates a broad peak ($2\theta\simeq21^\circ$) that originates from the epoxy glue used for embedding the PuFeAsO powder. The insert shows a zoom of the 42 - 60 degrees interval to highlight the excellent quality of the fit: $R_\mathrm{wp}=2.8$, $R_\mathrm{exp}=2.46$, $\mathrm{\chi^{2}}=1.29$.}
\end{figure}

The room temperature x-ray powder diffraction profile is presented in Fig.~\ref{xrd}.  Two impurity phases can be detected: PuAs and PuO$_2$ of $3\%$ and $6\%$ respectively. PuFeAsO is an isostructural analogue of the rare-earth iron oxypnictide family \textit{R}FeAsO, as can be seen by the successful refinement of the XRD data, shown in Fig. ~\ref{xrd}, to the ZrCuSiAs-type structure. To facilitate comparison, crystallographic parameters for \textit{M}FeAsO (\textit{M} = Np, Pu) as well as for \textit{R}FeAsO (\textit{R} = La, Tb) derived from Rietveld refinement of the structural model are given in Table I.

\begin{figure}[t]
\centering
\includegraphics[width=0.45\textwidth,bb=35 160 440 620,clip]{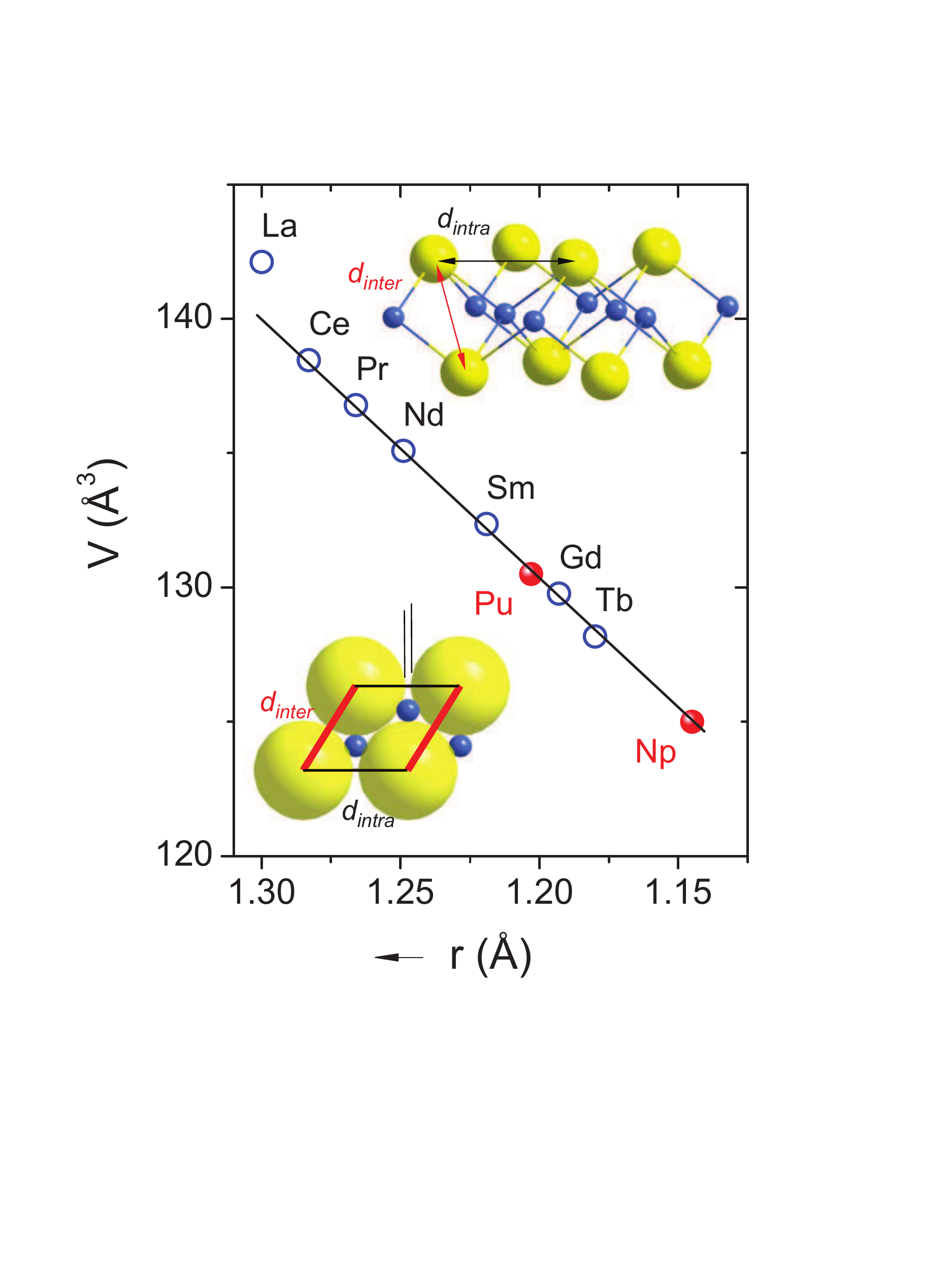}\caption
{\label{volume}(Color online) The unit cell volume for the rare-earth \textit{R}-1111 series vs. the ionic radius of the \textit{R} ion. The volume values have been calculated based on the data from ref. \onlinecite{Nitsche}, and the radii of the corresponding trivalent \textit{R} metal in octahedral coordination were found in ref. \onlinecite{shannon}. The red, solid points on the diagram represent trivalent Np and Pu, based on the unit cell volume obtained for NpFeAsO and PuFeAsO, respectively.
The upper inset shows the As (large yellow spheres) and Fe (small blue spheres) planes, highlighting two different As -- As distances: $d_\mathrm{intra}$ and $d_\mathrm{inter}$. These distances are also labelled on an alternative projection in the lower inset.}
\end{figure}

\begin{table}[b!]
\centering
\begin{tabular}{ccccc}
\hline\hline\\
\smallskip
& LaFeAsO & TbFeAsO & NpFeAsO & PuFeAsO \\
 \hline
\textit{a} (\AA) & 4.0367(1) & 3.9043(3) & 3.86369(4) & 3.91925(5) \\
\textit{c} (\AA) & 8.7218(4) & 8.408(1) & 8.36125(11) & 8.49792(17) \\
V (\AA$^{3}$) & 142.12(1) & 128.17(3) & 124.82(1) & 130.53(2) \\
\textit{z}(M) & 0.14141(5) & 0.13455(3) & 0.15477(14) & 0.14607(13) \\
\textit{z}(As) & 0.65138(9) & 0.66389(6) & 0.6709(4) & 0.6606(4) \\
\hline
d$_\mathrm{inter}$(\AA) & 3.888(1) & 3.901(1) & 3.954(4) & 3.890(4) \\
d$_\mathrm{Fe-As}$(\AA) & 2.4128(4) & 2.3895(3) & 2.403(2) & 2.388(2) \\
$\alpha$$_\mathrm{Fe-As-Fe}$ & $113.6^{\circ}$ & $109.6^{\circ}$ & $107.0^{\circ}$ & $110.3^{\circ}$ \\
\hline
R$_\mathrm{wp}$ & -- & -- & 5.38 & 2.8 \\
R$_\mathrm{exp}$ & -- & -- & 4.68 & 2.46 \\
$\chi^{2}$ & -- & -- & 1.32 & 1.29 \\
\hline \hline
\end{tabular}\caption{\label{table1}Lattice parameters, volume, and $d_\mathrm{inter}$ distance for LaFeAsO, TbFeAsO, NpFeAsO and PuFeAsO at room temperature. The data for LaFeAsO and TbFeAsO are taken from ref. \onlinecite{Nitsche}.}
\end{table}

Figure~\ref{volume} presents the unit cell volume for the rare-earth \textit{R}-1111 series as a function of the ionic radius $r$ of the $\textit{R}^{3+}$ ion. Assuming that the Np and Pu trivalent ions have the same crystallographic environment, we can deduce the ionic radii of Pu$^{3+}$ and Np$^{3+}$. The radii were found based on the calculated unit cell volume for PuFeAsO, NpFeAsO and the $V(r)$ trend in the \textit{R}FeAsO series, which in Figure~\ref{volume} is represented by a straight line through the data. Estimated radii are 1.20 and 1.14~\AA\ for Pu$^{3+}$ and Np$^{3+}$, respectively. The ionic radii values are larger than those reported by Shannon and Prewitt\cite{shannon} ($r(Pu^{3+}) = 1.00$~\AA\ and $r(Np^{3+}) = 1.04$~\AA), although in that case they were for coordination number $N = 6$. This discrepancy may be explained by the fact that increasing the coordination number (for the same oxidation state), causes an increase of the ionic radii, e.g. $r(Tb^{3+})$ is 0.923~\AA\ and 1.04~\AA, for $N = 6$ and $N = 8$, respectively\cite{shannon}. To our knowledge, the radii for eightfold coordinated trivalent neptunium and plutonium have not been previously reported.

In ref.~\onlinecite{Nitsche} Nitsche \emph{et al.} have successfully employed a hard-sphere model in order to understand the structural trends in the \textit{R}FeAsO system. The model reveals a geometrical limit for the structure and clearly explains why the \textit{R}FeAsO compounds with the heavier (and smaller) \textit{R} lanthanides can be obtained only via high pressure synthesis methods.
The basic arguments behind this hard sphere model are illustrated in the lower inset of Fig. ~\ref{volume}. This shows that the interplanar distance $d_\mathrm{inter}$ (thick red line) is equal to $2r_\mathrm{As}$, since in this direction the As ions (large yellow spheres) are in contact. This distance slightly increases from 3.889 \AA to 3.901 \AA, for LaFeAsO and TbFeAsO respectively.
Meanwhile, the intraplanar distance $d_\mathrm{intra}$ (thin black line) is along $[100]$, and equal to the $a$ lattice parameter. The hard sphere model reveals that the \textit{R}FeAsO structure is formed only for $d_\mathrm{intra} \geq d_\mathrm{inter}$. The geometrical limit can be seen by the disappearing of the gap between the As ions for $d_\mathrm{intra} = d_\mathrm{inter}$. This is the case for TbFeAsO, which lies at the border of stability and therefore DyFeAsO and HoFeAsO can be synthesized only by high-pressure methods\cite{Rodgers09, Bos2008}. The criterion is satisfied for PuFeAsO. However, for NpFeAsO $d_\mathrm{intra}$ is smaller than $d_\mathrm{inter}$, despite the fact that NpFeAsO may be synthesized without recourse to high pressure\cite{Klimczuk2012}, which means that the hard sphere model does not work for NpFeAsO. This is likely to be due to the less ionic character of the chemical bonds in NpFeAsO compared to PuFeAsO.

\begin{figure}[t]
\centering
\includegraphics[width=0.475\textwidth,clip]{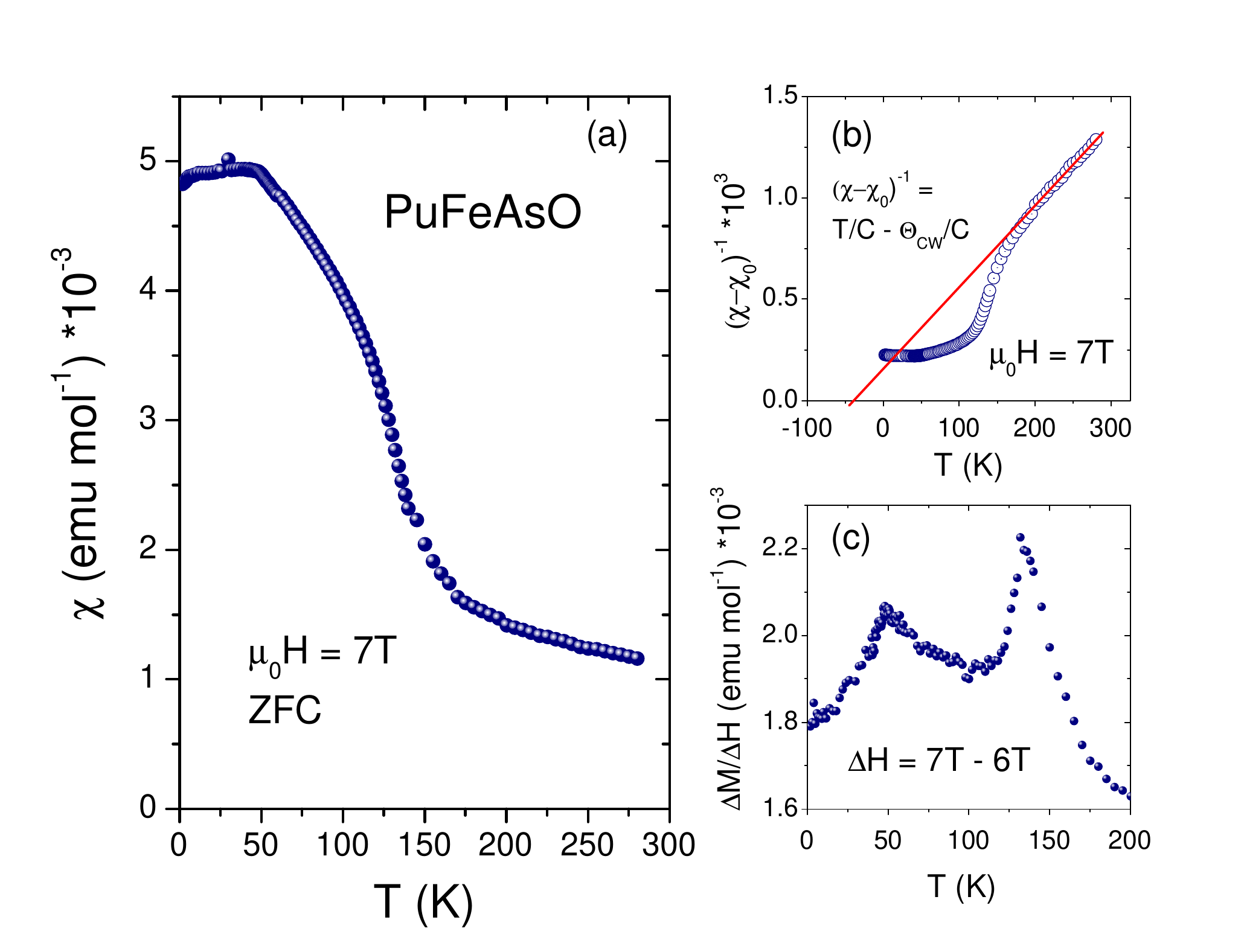}\caption
{\label{figure3}(Color online) Panel a):  zero-field-cooling magnetic susceptibility of PuFeAsO, corrected for the presence of PuAs and PuO$_2$ as described in the text. Panel b): inverse magnetic susceptibility and a Curie-Weiss fit to the high temperature part, yielding an effective moment of $1.4\pm0.2$~$\mu_B$. Panel c): the difference in magnetization curves $\Delta$M = M(T, H$_{2}$) - M(T,H$_{1}$) measured  with applied fields H$_{2}$ = 7~T and H$_{1}$ = 6~T, showing two magnetic anomalies in our PuFeAsO sample.}
\end{figure}

Panel a) of Fig.~\ref{figure3} shows the $\chi(T)$ curve measured after zero-field-cooling (ZFC) with an applied magnetic field of $7$~T. The rapid increase of $\chi(T)$ below $150$~K is caused by the onset of ferromagnetic order in the PuAs impurity, characterized by a Curie temperature of $129$~K \cite{Blaise-PuAs}. At $50$~K, $\chi(T)$ reaches a maximum and then slightly decreases with decreasing temperature, which suggests the presence of a magnetic transition at this temperature. The inverse susceptibility curve is plotted in panel b) of Figure~\ref{figure3}. Above $200$~K the magnetic susceptibility can be fitted by a modified Curie-Weiss law to obtain a Curie-Weiss temperature of $-42\pm9$~K (suggestive of the presence of antiferromagnetic fluctuations), a temperature independent susceptibility $\chi_{0} = (3.8\pm1.2)\times10^{-4}$ emu mol$^{-1}$, and an effective paramagnetic moment of $1.4\pm0.2~\mu_B$. The above values have been obtained by fitting the $(\chi-\chi_{0})^{-1}$ curve to a straight line, taking into account the contribution of the impurity phases, PuO$_2$ (temperature independent paramagnetism (ref. \onlinecite{Raphael68})) and PuAs. For the latter compound, only the Curie-Weiss paramagnetic signal above $200$~K was subtracted, which was obtained based on the fitting parameters reported in ref. \onlinecite{Blaise-PuAs}.


The obtained $\mu_\mathrm{eff}$ of $1.4\pm0.2$~$\mu_B$ for PuFeAsO, is almost twice as large as the LS-coupling value of 0.85 $\mu_{B}$ expected for a Pu$^{3+}$ ion with a Russell-Saunders $^{6}H_{5/2}$ ground state. This discrepancy might be due to the additional magnetism of the Fe ion in PuFeAsO, as suggested by the band structure calculations described below. However, assuming intermediate coupling, as we did for NpFeAsO\cite{Klimczuk2012}, and zero crystal-field splitting, the calculated effective moment would be approximately 1.4 $\mu_{B}$ (ref. \onlinecite{Le2010}), in very good agreement with our measurement.

Panel c) of Fig.~\ref{figure3} shows the difference ($\Delta M$) between the temperature-dependent magnetization curves measured in an applied field of $7$~T and $6$~T, divided by $\Delta H = 1$~T. This procedure is used in order to subtract the ferromagnetic signal of the unreacted ferromagnetic impurities\cite{Klimczuk04}.  The susceptibility ($\Delta M$/$\Delta H$) derived in such a fashion shows two anomalies. By analogy to results from NpFeAsO \cite{Klimczuk2012}, the anomaly at 50~K is attributed to antiferromagnetic ordering of PuFeAsO, whereas the anomaly at higher temperature is caused by the PuAs ferromagnetic transition at $T = 129$~K.\cite{Blaise-PuAs}

\begin{figure} [t]
\centering
\includegraphics[width=0.45\textwidth,clip]{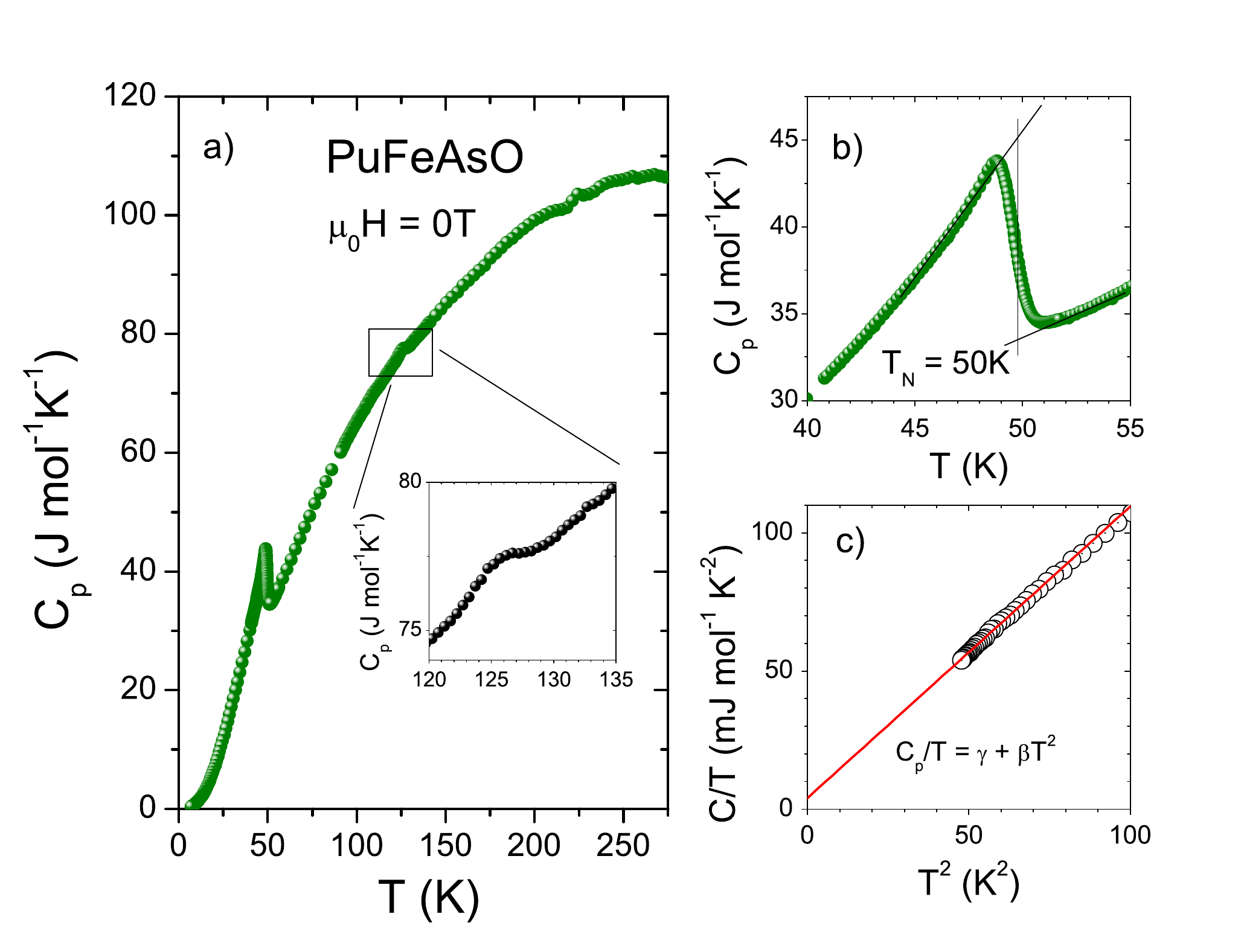}\caption
{\label{figure4}(Color online) Panel a): temperature dependence of the heat capacity $C_{p}$ of PuFeAsO and a blow-up of the temperature interval around $126$~K, where the small anomaly due to the ferromagnetic transition of PuAs occurs. Panel b): a sharp lambda anomaly at $T_N\sim50$~K reveals the stabilization of antiferromagnetic order in PuFeAsO. Panel c): $C_{p}/T$ as a function of $T^{2}$ and a fit to $C_{p}/T=\gamma+\beta T^2$.}
\end{figure}

The temperature dependence  of the heat capacity ($C_{p}$) is shown in Fig.~\ref{figure4}a). The raw data have been corrected for the additional heat capacity of the stycast and PuO$_2$ impurity phase by scaling the data reported in ref.~\onlinecite{Flotow1976}. A small anomaly is seen at $126$~K corresponding to the ferromagnetic transition temperature in the PuAs impurity phase, but the large lambda anomaly at $50$~K confirms the bulk nature of the transition in PuFeAsO and this is zoomed on in Fig.~\ref{figure4}b in order to define the N\'{e}el temperature of PuFeAsO.

A straight line through the data in Fig.~\ref{figure4}c shows the fit of $C_{p}/T=\gamma+\beta T^2$ in the range $7 < T < 10$ K. The fit reveals an electronic heat capacity $\gamma=4$~mJ~K$^{-2}$~mol$^{-1}$, and a Debye temperature $\theta_D=194$~K (where $\beta=12\pi^4Nk_B/5\theta^3_D$). Although the experimental data were fitted only up to 10K, which is five times lower than $T_{N}$, we are aware of the magnetic contribution ($C_{mag}$) to the specific heat. Subtracting $C_{mag}$ should not change the value of $\gamma$, but might slightly change $\beta$, and consequently the Debye temperature. Whilst the electronic heat capacity of PuFeAsO is an order of magnitude smaller than that of NpFeAsO; their Debye temperatures are not very different, as expected from the similarities in the lattice parameters and the very small change in atomic masses.

To further investigate the electronic and magnetic character of PuFeAsO, we have performed first-principles local-spin-density approximation (LSDA) and LSDA plus Coulomb-U (LSDA+U) calculations assuming the tetragonal ZrCuSiAs crystal structure with crystallographic parameters determined experimentally at room temperature. Gaining inspiration from neutron scattering data for structural analogues, we assumed that the magnetic and crystallographic unit cells coincide. Furthermore, we assumed that the magnetic moment is aligned along the $c$-axis, as it is the case for NpFeAsO\cite{Klimczuk2012}, and considered non-magnetic (NM), ferromagnetic (FM), and anti-ferromagnetic (AF) arrangements for Pu atoms.

We used the in-house implementation of the full-potential linearized augmented plane wave (FP-LAPW) method \cite{shick99}. This FP-LAPW version includes all relativistic effects (scalar-relativistic and spin-orbit coupling), and relativistic implementation of the rotationally invariant LSDA+U \cite{shick01}. In the FP-LAPW calculations we set the radii of the atomic spheres to 2.75~a.u.~(Pu), 2.2~a.u.~(Fe, As), and 1.6~a.u.~(O). The parameter $R_{Np} \times K_{\text{max}}=9.625$ determined the size of the basis set and the Brillouin zone (BZ) sampling was performed with 405 $k$~points.  For the plutonium $f$~shell, Slater integrals of $F_0 = 4.00$~eV, $F_2=7.76$~eV, $F_4=5.05$~eV and $F_6= 3.70$~eV were selected to specify the Coulomb interaction~\cite{KMoore2009}. They correspond to commonly accepted values for Coulomb $U = 4$~eV and exchange $J = 0.64$~eV parameters. The so-called fully-localized-limit (FLL) flavor for LSDA+U double-counting correction is used in the calculations. Following previous analysis for the NpFeAsO case~\cite{Klimczuk2012}, the choice of the FLL double counting is expected to be most appropriate for the PuFeAsO
material.

\begin{figure}[t]
\centering
    \includegraphics[angle=0,width=0.45\textwidth,clip]{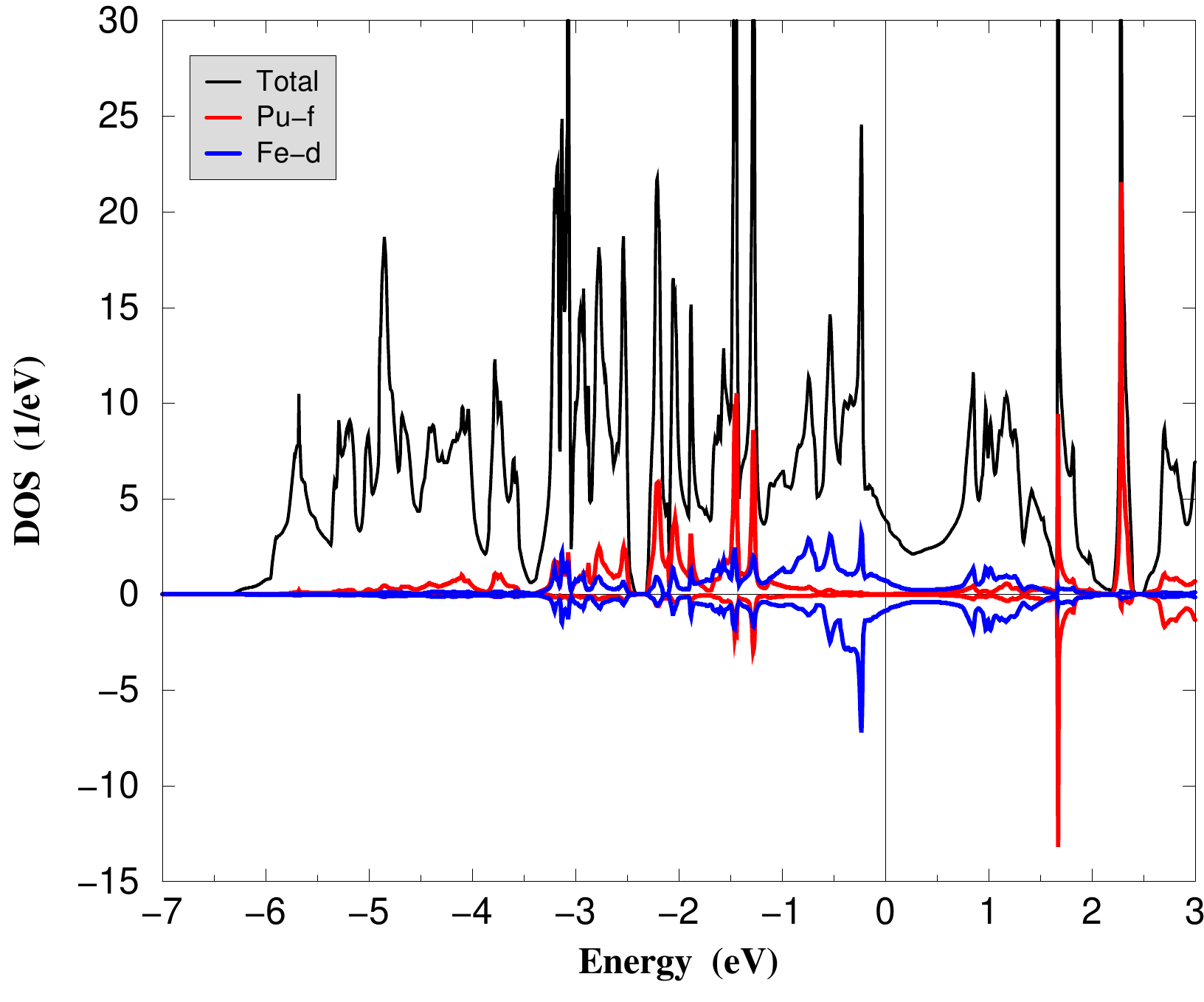}
    \includegraphics[width=0.45\textwidth,clip]{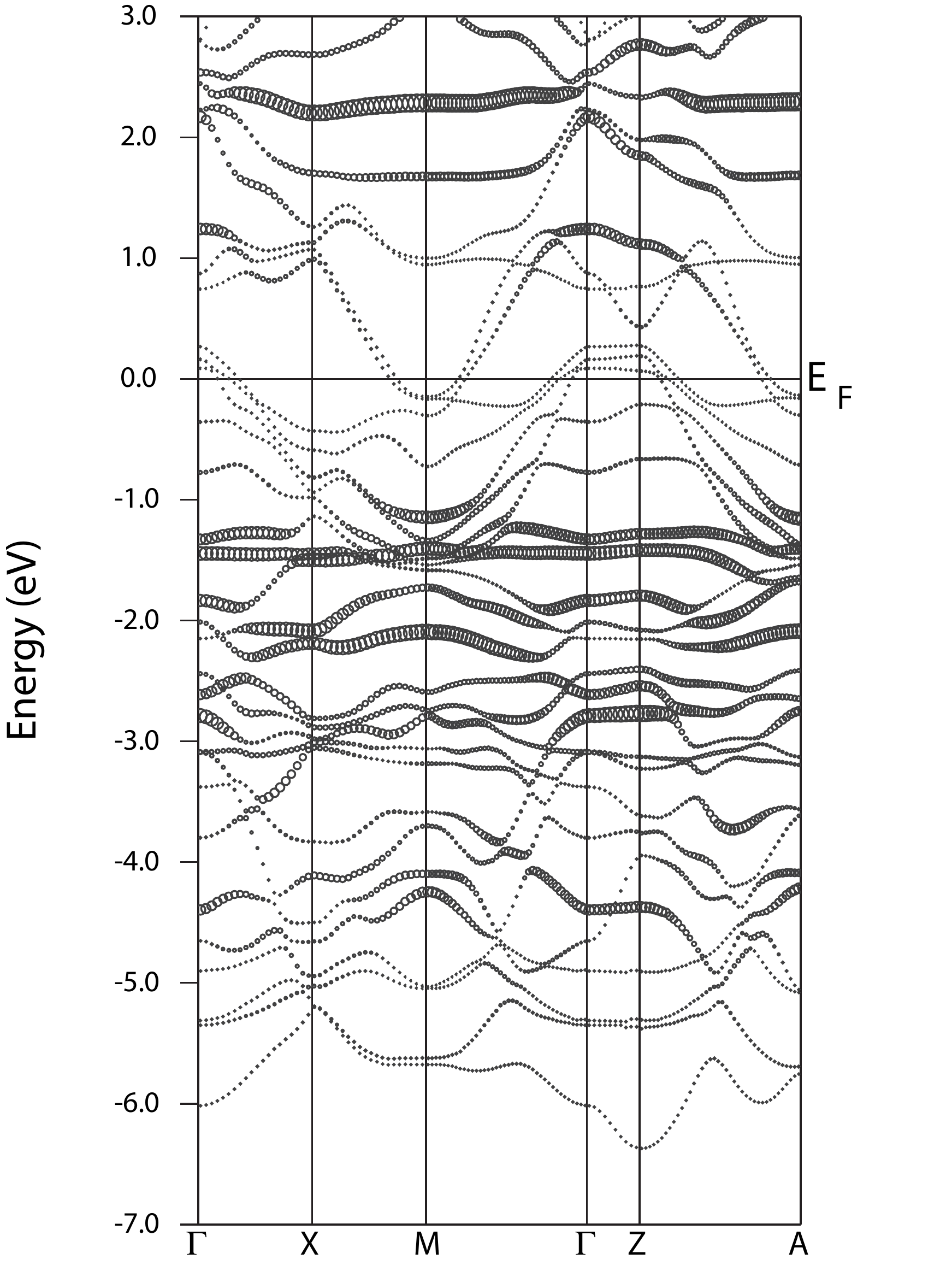}
    \caption
{(Color online) Electronic structure calculations for PuFeAsO: (top) the relativistic spin- and orbital-resolved density of states (DOS) per unit cell, including the partial f-DOS on Pu atom and d-DOS on Fe atom; (bottom) Band structure of PuFeAsO with f-weight fatbands.} \label{fig:DOS}
\end{figure}

In LSDA, we found that spin-polarization decreases the total energy with respect to the non-magnetic solution by 0.877 eV/ per formula unit (f.u.) for the FM-solution, and by 0.776 eV/f.u. for the AF-solution, suggesting an FM-ordered ground state. However, when we included the Coulomb-$U$ and exchange-$J$ using the LSDA+U method, the FM-solution is higher in the total energy than the AF-solution by 4.4 meV/f.u in agreement with the bulk measurements. It should be noted that the total energy difference between the FM and the AF solutions in PuFeAsO is substantially smaller than in NpFeAsO~\cite{Klimczuk2012}.

The  spin $m_S$, orbital $m_L$, and total $m_J$ magnetic moments for the AF calculations with LSDA and LSDA+U are shown in Table~\ref{tab:2}. The staggered local magnetic moments, which are due to magnetic polarization of the Pu $f$-shell, are formed at the two Pu atoms in the unit cell. The small magnitude of the ordered local magnetic moment $m_J$ is due to partial compensation of its spin $m_S$ and orbital $m_L$ components.

The exchange splitting at the Pu atoms forces the induced staggered spin moments at the Fe atoms in a checker-board AF arrangement. Contrary to the NpFeAsO case, where the Fe atom moments are fairly small, they reach $\sim$0.5 $\mu_\mathrm{B}$ in PuFeAsO. Such moments on the iron could at least in principle be detected in future neutron powder diffraction or M\"{o}ssbauer measurements.

\begin{table}
  \begin{tabular}{cccccccccc}
         \hline
 &\multicolumn{4}{c}{\bf  LSDA} &&\multicolumn{4}{c}{\bf LSDA+U} \\
            & Pu & Fe & As & O & & Pu & Fe & As & O \\
           \hline
 $m_S$&4.30 & 0.62 & 0.03 & 0.00  & &3.65 &0.51 & 0.01 & 0.00  \\
 $m_L$&-2.87& 0.03 & 0.00  & 0.00 & &-4.06&0.03 & 0.00 & 0.00  \\
 $m_J$&1.43 & 0.65 & 0.03  &0.00  & &-0.41&0.54 & 0.01 & 0.00  \\
           \hline
  \end{tabular}
    \caption{Spin $m_S$, orbital $m_L$, and total $m_J$ magnetic moments ($\mu_\mathrm{B}$), for one (of two) Pu, Fe, As, and O atoms in anti-ferromagnetic PuFeAsO, resulting from calculations with LSDA and LSDA+U=4 eV with FLL-double-counting flavor.}
    \label{tab:2}
\end{table}

The LSDA+U total and partial (atom, spin and orbital-resolved) density of states (DOS)  are shown in Fig.~\ref{fig:DOS}. The total DOS at $E_F$ of 4.0 states eV$^{-1}$ corresponds to a non-interacting value of the Sommerfeld coefficient $\gamma$= 4.7 mJ~K$^{-2}$~mol$^{-1}$, which agrees well with the experimental value of 4 mJ~K$^{-2}$~mol$^{-1}$. The DOS near the Fermi energy ($E_F$) has mostly Fe-$d$ character (see Fig.~\ref{fig:DOS}), whilst the As-$p$ and
O-$p$ states are mostly located at 2-7 eV energy interval below $E_F$. The Pu-$f$ states are split by the exchange interaction. The f-character, which is shown by fatbands in Fig.~\ref{fig:DOS}, is shifted away from the Fermi level.

\begin{figure}
\centering
    \includegraphics[angle=-90,width=0.5\textwidth,clip]{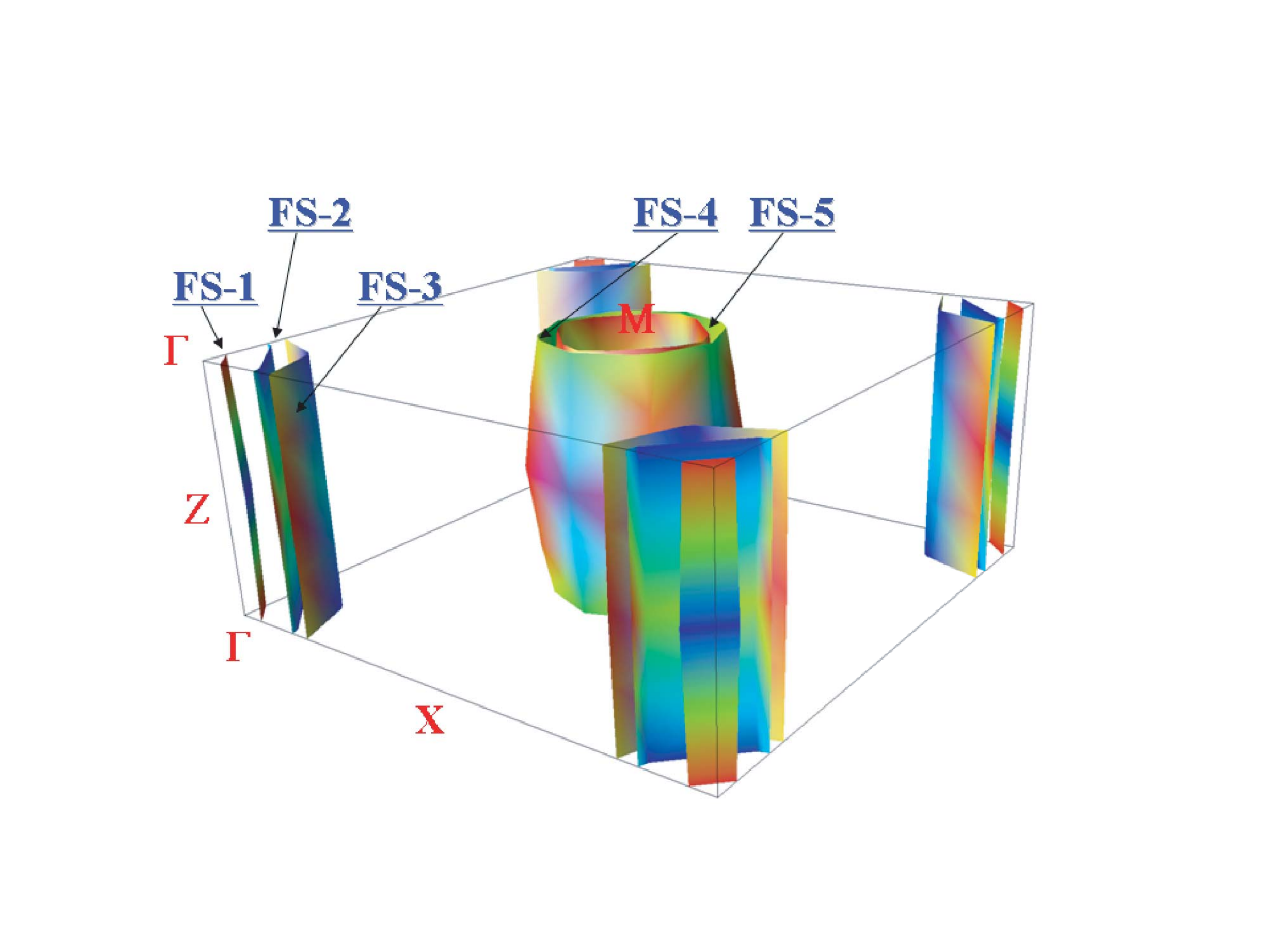}
\caption{(Color online) Fermi surface of PuFeAsO exhibiting five double-degenerate sheets with pronounced two-dimensional character, three hole-like (FS-1, 2, 3) and two electron-like (FS-4, 5). The shade of colors encodes the energy gradient.} \label{fig:fermisurface}
\end{figure}

The Fermi surface (FS)  is shown in Fig.~\ref{fig:fermisurface}, and  consists of five sheets, each of them doubly degenerate. Examination of the band structure shows that FS sheets $1-3$ are hole-like, and centered at the $\Gamma$-point. FS-4 and FS-5 are electron-like and centered at the M-point. Note the fairly two dimensional character of  the FS.

Finally, we calculate the electronic structure of PuFeAsO in the paramagnetic phase. We use a dynamical mean-field-like approach combining the local density approximation (LDA) with the exact diagonalization (ED)~\cite{kolorenc2012} of the single impurity Anderson model which include the energy dependent self-energy $\Sigma$. These LDA+ED calculations also include self-consistency over the charge density in the FP-LAPW basis employing the procedure of ref.~\onlinecite{shick09}. In these calculations, we keep the same Slater integrals as above, and set the inverse temperature $\beta$=235 eV$^{-1}$ ($T$=50 K).

\begin{figure}
\centerline{\includegraphics[angle=0,width=0.45\textwidth,clip]{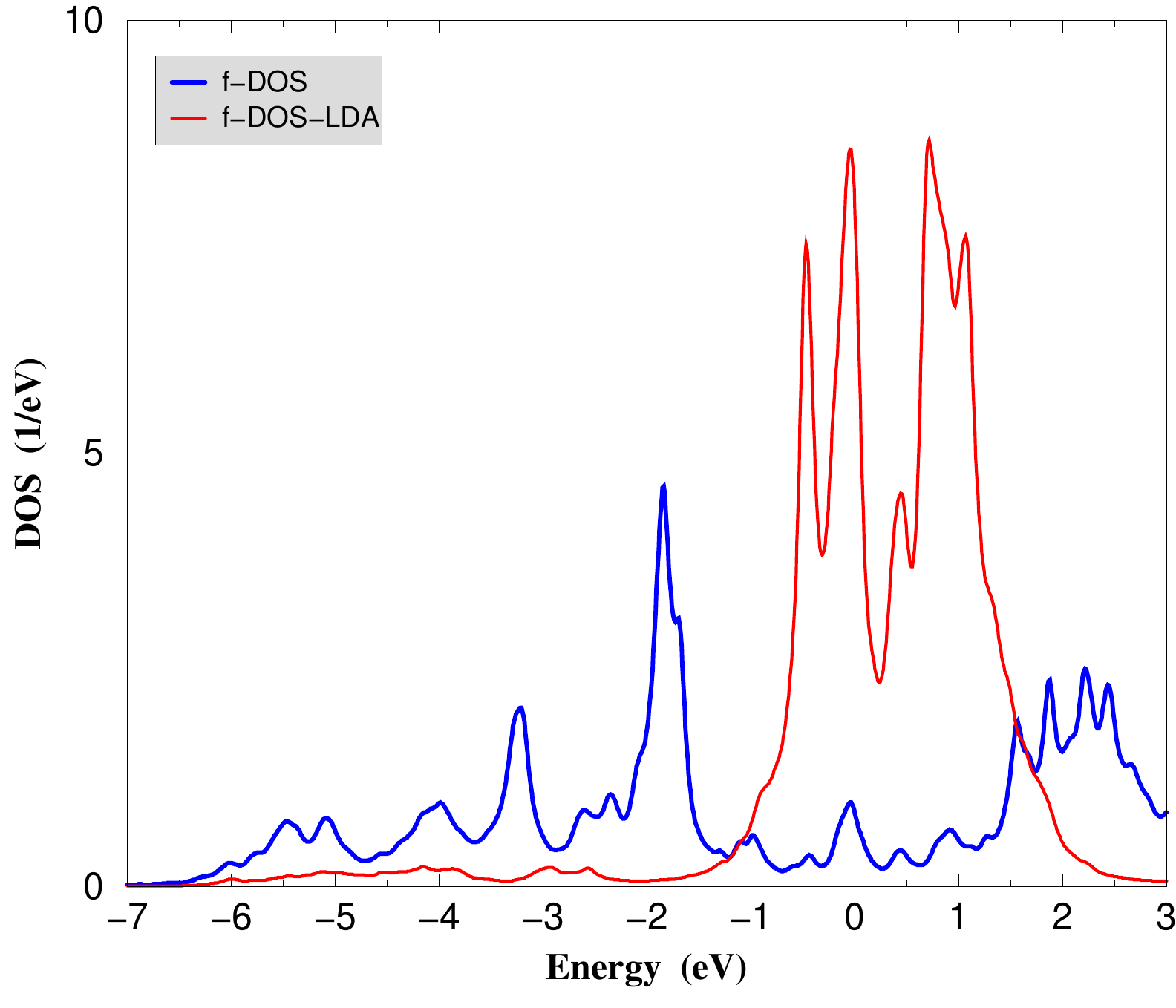}}
\caption{(Color online)  The $f$-electron spectral density of states (DOS) for the Pu atom in PuFeAsO from paramagnetic LDA+ED calculations in comparison with LDA.} \label{fig:DOSED}
\end{figure}

The resulting $f$-orbital DOS is shown in Fig.~\ref{fig:DOSED}. It is seen that a small quasi-particle peak is formed at $E_F$ due to hybridization between f- and non-f-states. Most of the f-character weight is shifted into Hubbard bands, at variance with LDA. Often, LDA (or its semi-local analog generalized gradient approximation (GGA)) is used to model a paramagnetic state of f-electron systems. Our calculations show that LDA/GGA are not appropriate, and that electron correlation effects play an essential role in PuFeAsO.

The corresponding ground state for the Pu atom in PuFeAsO is a composite cluster of electrons formed by the $f$-shell and the hybridization bath, containing $\langle n_f+n_{bath} \rangle$ = 10.73 particles, with $\langle n_f \rangle$ = 5.01. A similar calculation performed for the PuCoGa$_{5}$ superconductor yields $\langle n_f \rangle$ = 5.30 and  $\langle n_f+n_{bath} \rangle$ = 14, corresponding to a singlet ground state for the composite cluster \cite{daghero12}. This suggests that the $f$ electrons in PuFeAsO are less hybridized with the conduction band than in PuCoGa$_{5}$ and are close to a static integer valence.

The expectation values $\langle \hat X^2 \rangle = X(X+1)$ ($X$ = $S$, $L$, $J$) calculated for the $f$ shell in PuFeAsO give $S^{(5f)}$ = 2.21, $L^{(5f)}$ = 4.63 and $J^{(5f)}$ = 2.52, for the spin, orbital and total moments, respectively. The individual components of the moments, $\langle \hat S^{(5f)}_z \rangle$ and $\langle \hat L^{(5f)}_z \rangle$, however vanish, so that the spin-orbital symmetry is preserved and neither spin nor orbital polarization is induced in the absence of the external magnetic field. Subsequently, the effective magnetic moment $\mu_{eff} = g_J \sqrt{J(J+1)}= 0.25 \; \mu_B$ is evaluated, where the g-factor $g_J = (L-2S)/J$ is used. This is smaller than the experimental moment derived from the magnetic susceptibility measurements. This difference between the theory and experiment can be interpreted as a consequence of the non-zero local moment of the Fe atom.

\section{Conclusions}

In summary, we have succeeded in synthesizing a plutonium analog of the  iron-based oxypnictide superconductors \textit{R}FeAsO. Powder x-ray diffraction at room temperature shows that PuFeAsO has the same tetragonal layered structure exhibited by LaFeAsO, with a unit cell volume close to GdFeAsO. A comparison of the relevant crystallographic parameters suggests that the stability of PuFeAsO can be well described by the hard-sphere model proposed for understanding the structural trends in the \textit{R}-1111 system \cite{Nitsche}, implying a Pu$^{3+}$ valence state with a well localized $5f$ electronic configuration. This is not the case for NpFeAsO, where the chemical bonds seem to have a more covalent or transition-metal-like character.

Magnetic susceptibility and specific heat measurements have been used to characterize the macroscopic physical properties of PuFeAsO. As the samples were prepared using $^{239}$Pu, large self-heating effects associated to $\alpha$ decay limited the lowest attainable temperature in specific heat measurements to 7 K. Down to this temperature we observe only one anomaly related to PuFeAsO, suggesting the stabilization of antiferromagnetic order below $T_N=50$~K. No further anomaly is observed between $\sim$2 and 7 K in the magnetic susceptibility curve. A similar behavior is exhibited by NpFeAsO, where the magnetic order at the Np site below 57 K has been confirmed by neutron diffraction \cite{Klimczuk2012}. A $T_N$ in PuFeAsO smaller than in the Np analog implies a smaller exchange interaction. On the other hand, electronic structure calculations suggest that the induced staggered magnetic moment on the Fe sublattice of PuFeAsO is comparable to that carried by the Pu atoms, whereas in the case of NpFeAsO the induced Fe magnetic moment is calculated to be one order of magnitude smaller than the Np moment. The ground state of the Pu atoms in the paramagnetic phase has been calculated by a method combining the local density approximation with an exact diagonalization of the single impurity Anderson model, in order to take into account the full structure of the $f$-orbital multiplet. The results indicate an occupation number of the Pu $5f$ shell very close to 5, corresponding to a stable Pu$^{3+}$ configuration, in contrast to $\delta$-Pu and PuCoGa$_{5}$ for which an intermediate valence configuration has been suggested \cite{daghero12}.

The high temperature anomalies that are always present in the parent \textit{R}FeAsO compounds are not observed in PuFeAsO, and so we would suggest that, similar to NpFeAsO, spin density wave condensation and structural phase transition are not present in PuFeAsO. Continuing the trend from Np to Pu and on to Am, one might imagine that the increased degree of localisation of the Am $5f$ electrons might be more compatible with a magnetic structure on the iron independent of that on the $f$ electron ion, such that the SDW and related structural transition might be present.

\section{Acknowledgements}
 Pu metal required for the fabrication of the compound was made available through a loan agreement between Lawrence Livermore National Laboratory and ITU, in the framework of a collaboration involving Lawrence Livermore National Laboratory, Los Alamos National Laboratory, and the US Department of Energy. Participation in the European Community Transnational Access to Research Infrastructures Action of the "Structuring the European Research Area" specific programme contract $\mathrm{RITA CT 200C026176}$, is acknowledged. The support from Czech Republic Grants GACR P204/10/0330, and GAAV IAA100100912 is thankfully acknowledged.

\bibliographystyle{apsrev}
\bibliography{pufeaso}

\begin{thebibliography}{22}
\expandafter\ifx\csname natexlab\endcsname\relax\def\natexlab#1{#1}\fi
\expandafter\ifx\csname bibnamefont\endcsname\relax
  \def\bibnamefont#1{#1}\fi
\expandafter\ifx\csname bibfnamefont\endcsname\relax
  \def\bibfnamefont#1{#1}\fi
\expandafter\ifx\csname citenamefont\endcsname\relax
  \def\citenamefont#1{#1}\fi
\expandafter\ifx\csname url\endcsname\relax
  \def\url#1{\texttt{#1}}\fi
\expandafter\ifx\csname urlprefix\endcsname\relax\def\urlprefix{URL }\fi
\providecommand{\bibinfo}[2]{#2}
\providecommand{\eprint}[2][]{\url{#2}}

\bibitem[{\citenamefont{Zwicknagl and Fulde}(2003)}]{Zwicknagl}
\bibinfo{author}{\bibfnamefont{G.}~\bibnamefont{Zwicknagl}} \bibnamefont{and}
  \bibinfo{author}{\bibfnamefont{P.}~\bibnamefont{Fulde}},
  \bibinfo{journal}{Journal of Physics: Condensed Matter}
  \textbf{\bibinfo{volume}{15}}, \bibinfo{pages}{S1911} (\bibinfo{year}{2003}).

\bibitem[{\citenamefont{Santini et~al.}(2009)\citenamefont{Santini, Carretta,
  Amoretti, Caciuffo, Magnani, and Lander}}]{santini09}
\bibinfo{author}{\bibfnamefont{P.}~\bibnamefont{Santini}},
  \bibinfo{author}{\bibfnamefont{S.}~\bibnamefont{Carretta}},
  \bibinfo{author}{\bibfnamefont{G.}~\bibnamefont{Amoretti}},
  \bibinfo{author}{\bibfnamefont{R.}~\bibnamefont{Caciuffo}},
  \bibinfo{author}{\bibfnamefont{N.}~\bibnamefont{Magnani}}, \bibnamefont{and}
  \bibinfo{author}{\bibfnamefont{G.~H.} \bibnamefont{Lander}},
  \bibinfo{journal}{Rev. Mod. Phys.} \textbf{\bibinfo{volume}{81}},
  \bibinfo{pages}{807} (\bibinfo{year}{2009}).

\bibitem[{\citenamefont{Hecker}(2003)}]{Hecker2003}
\bibinfo{author}{\bibfnamefont{S.}~\bibnamefont{Hecker}},
  \bibinfo{journal}{JOM} \textbf{\bibinfo{volume}{55}}, \bibinfo{pages}{13}
  (\bibinfo{year}{2003}).

\bibitem[{\citenamefont{Joyce et~al.}(2003)\citenamefont{Joyce, Wills,
  Durakiewicz, Butterfield, Guziewicz, Sarrao, Morales, Arko, and
  Eriksson}}]{Joyce2003}
\bibinfo{author}{\bibfnamefont{J.~J.} \bibnamefont{Joyce}},
  \bibinfo{author}{\bibfnamefont{J.~M.} \bibnamefont{Wills}},
  \bibinfo{author}{\bibfnamefont{T.}~\bibnamefont{Durakiewicz}},
  \bibinfo{author}{\bibfnamefont{M.~T.} \bibnamefont{Butterfield}},
  \bibinfo{author}{\bibfnamefont{E.}~\bibnamefont{Guziewicz}},
  \bibinfo{author}{\bibfnamefont{J.~L.} \bibnamefont{Sarrao}},
  \bibinfo{author}{\bibfnamefont{L.~A.} \bibnamefont{Morales}},
  \bibinfo{author}{\bibfnamefont{A.~J.} \bibnamefont{Arko}}, \bibnamefont{and}
  \bibinfo{author}{\bibfnamefont{O.}~\bibnamefont{Eriksson}},
  \bibinfo{journal}{Phys. Rev. Lett.} \textbf{\bibinfo{volume}{91}},
  \bibinfo{pages}{176401} (\bibinfo{year}{2003}).

\bibitem[{\citenamefont{Zhu et~al.}(2007)\citenamefont{Zhu, McMahan, Jones,
  Durakiewicz, Joyce, Wills, and Albers}}]{Zhu2007}
\bibinfo{author}{\bibfnamefont{J.-X.} \bibnamefont{Zhu}},
  \bibinfo{author}{\bibfnamefont{A.~K.} \bibnamefont{McMahan}},
  \bibinfo{author}{\bibfnamefont{M.~D.} \bibnamefont{Jones}},
  \bibinfo{author}{\bibfnamefont{T.}~\bibnamefont{Durakiewicz}},
  \bibinfo{author}{\bibfnamefont{J.~J.} \bibnamefont{Joyce}},
  \bibinfo{author}{\bibfnamefont{J.~M.} \bibnamefont{Wills}}, \bibnamefont{and}
  \bibinfo{author}{\bibfnamefont{R.~C.} \bibnamefont{Albers}},
  \bibinfo{journal}{Phys. Rev. B} \textbf{\bibinfo{volume}{76}},
  \bibinfo{pages}{245118} (\bibinfo{year}{2007}).

\bibitem[{\citenamefont{Kamihara et~al.}(2006)\citenamefont{Kamihara,
  Hiramatsu, Hirano, Kawamura, Yanagi, Kamiya, and Hosono}}]{Kamihara2006}
\bibinfo{author}{\bibfnamefont{Y.}~\bibnamefont{Kamihara}},
  \bibinfo{author}{\bibfnamefont{H.}~\bibnamefont{Hiramatsu}},
  \bibinfo{author}{\bibfnamefont{M.}~\bibnamefont{Hirano}},
  \bibinfo{author}{\bibfnamefont{R.}~\bibnamefont{Kawamura}},
  \bibinfo{author}{\bibfnamefont{H.}~\bibnamefont{Yanagi}},
  \bibinfo{author}{\bibfnamefont{T.}~\bibnamefont{Kamiya}}, \bibnamefont{and}
  \bibinfo{author}{\bibfnamefont{H.}~\bibnamefont{Hosono}},
  \bibinfo{journal}{J. Am. Chem. Soc.} \textbf{\bibinfo{volume}{128}},
  \bibinfo{pages}{10012} (\bibinfo{year}{2006}).

\bibitem[{\citenamefont{Klimczuk et~al.}(2012)\citenamefont{Klimczuk, Walker,
  Springell, Shick, Hill, Gaczy\ifmmode~\acute{n}\else \'{n}\fi{}ski, Gofryk,
  Kimber, Ritter, Colineau et~al.}}]{Klimczuk2012}
\bibinfo{author}{\bibfnamefont{T.}~\bibnamefont{Klimczuk}},
  \bibinfo{author}{\bibfnamefont{H.~C.} \bibnamefont{Walker}},
  \bibinfo{author}{\bibfnamefont{R.}~\bibnamefont{Springell}},
  \bibinfo{author}{\bibfnamefont{A.~B.} \bibnamefont{Shick}},
  \bibinfo{author}{\bibfnamefont{A.~H.} \bibnamefont{Hill}},
  \bibinfo{author}{\bibfnamefont{P.}~\bibnamefont{Gaczy\ifmmode~\acute{n}\else
  \'{n}\fi{}ski}}, \bibinfo{author}{\bibfnamefont{K.}~\bibnamefont{Gofryk}},
  \bibinfo{author}{\bibfnamefont{S.~A.~J.} \bibnamefont{Kimber}},
  \bibinfo{author}{\bibfnamefont{C.}~\bibnamefont{Ritter}},
  \bibinfo{author}{\bibfnamefont{E.}~\bibnamefont{Colineau}},
  \bibnamefont{et~al.}, \bibinfo{journal}{Phys. Rev. B}
  \textbf{\bibinfo{volume}{85}}, \bibinfo{pages}{174506}
  (\bibinfo{year}{2012}).

\bibitem[{\citenamefont{Flotow et~al.}(1976)\citenamefont{Flotow, Osborne,
  Fried, and Malm}}]{Flotow1976}
\bibinfo{author}{\bibfnamefont{H.~E.} \bibnamefont{Flotow}},
  \bibinfo{author}{\bibfnamefont{D.~W.} \bibnamefont{Osborne}},
  \bibinfo{author}{\bibfnamefont{S.~M.} \bibnamefont{Fried}}, \bibnamefont{and}
  \bibinfo{author}{\bibfnamefont{J.~G.} \bibnamefont{Malm}},
  \bibinfo{journal}{J. Chem. Phys.} \textbf{\bibinfo{volume}{65}},
  \bibinfo{pages}{1124} (\bibinfo{year}{1976}).

\bibitem[{\citenamefont{Nitsche et~al.}(2010)\citenamefont{Nitsche, Jesche,
  Hieckmann, Doert, and Ruck}}]{Nitsche}
\bibinfo{author}{\bibfnamefont{F.}~\bibnamefont{Nitsche}},
  \bibinfo{author}{\bibfnamefont{A.}~\bibnamefont{Jesche}},
  \bibinfo{author}{\bibfnamefont{E.}~\bibnamefont{Hieckmann}},
  \bibinfo{author}{\bibfnamefont{T.}~\bibnamefont{Doert}}, \bibnamefont{and}
  \bibinfo{author}{\bibfnamefont{M.}~\bibnamefont{Ruck}},
  \bibinfo{journal}{Phys. Rev. B} \textbf{\bibinfo{volume}{82}},
  \bibinfo{pages}{134514} (\bibinfo{year}{2010}).

\bibitem[{\citenamefont{Shannon and Prewitt}(1969)}]{shannon}
\bibinfo{author}{\bibfnamefont{R.~D.} \bibnamefont{Shannon}} \bibnamefont{and}
  \bibinfo{author}{\bibfnamefont{C.~T.} \bibnamefont{Prewitt}},
  \bibinfo{journal}{Acta Crystallogr. B} \textbf{\bibinfo{volume}{25}},
  \bibinfo{pages}{925} (\bibinfo{year}{1969}).

\bibitem[{\citenamefont{Rodgers et~al.}(2009)\citenamefont{Rodgers, Penny,
  Marcinkova, Bos, Sokolov, Kusmartseva, Huxley, and Attfield}}]{Rodgers09}
\bibinfo{author}{\bibfnamefont{J.~A.} \bibnamefont{Rodgers}},
  \bibinfo{author}{\bibfnamefont{G.~B.~S.} \bibnamefont{Penny}},
  \bibinfo{author}{\bibfnamefont{A.}~\bibnamefont{Marcinkova}},
  \bibinfo{author}{\bibfnamefont{J.-W.~G.} \bibnamefont{Bos}},
  \bibinfo{author}{\bibfnamefont{D.~A.} \bibnamefont{Sokolov}},
  \bibinfo{author}{\bibfnamefont{A.}~\bibnamefont{Kusmartseva}},
  \bibinfo{author}{\bibfnamefont{A.~D.} \bibnamefont{Huxley}},
  \bibnamefont{and} \bibinfo{author}{\bibfnamefont{J.~P.}
  \bibnamefont{Attfield}}, \bibinfo{journal}{Phys. Rev. B}
  \textbf{\bibinfo{volume}{80}}, \bibinfo{pages}{052508}
  (\bibinfo{year}{2009}).

\bibitem[{\citenamefont{Bos et~al.}(2008)\citenamefont{Bos, Penny, Rodgers,
  Sokolov, Huxley, and Attfield}}]{Bos2008}
\bibinfo{author}{\bibfnamefont{J.-W.~G.} \bibnamefont{Bos}},
  \bibinfo{author}{\bibfnamefont{G.~B.~S.} \bibnamefont{Penny}},
  \bibinfo{author}{\bibfnamefont{J.~A.} \bibnamefont{Rodgers}},
  \bibinfo{author}{\bibfnamefont{D.~A.} \bibnamefont{Sokolov}},
  \bibinfo{author}{\bibfnamefont{A.~D.} \bibnamefont{Huxley}},
  \bibnamefont{and} \bibinfo{author}{\bibfnamefont{J.~P.}
  \bibnamefont{Attfield}}, \bibinfo{journal}{Chem. Commun.} pp.
  \bibinfo{pages}{3634--3635} (\bibinfo{year}{2008}).

\bibitem[{\citenamefont{Blaise et~al.}(1973)\citenamefont{Blaise, Fournier, and
  Salmon}}]{Blaise-PuAs}
\bibinfo{author}{\bibfnamefont{A.}~\bibnamefont{Blaise}},
  \bibinfo{author}{\bibfnamefont{J.~M.} \bibnamefont{Fournier}},
  \bibnamefont{and} \bibinfo{author}{\bibfnamefont{P.}~\bibnamefont{Salmon}},
  \bibinfo{journal}{Solid State Commun.} \textbf{\bibinfo{volume}{13}},
  \bibinfo{pages}{555} (\bibinfo{year}{1973}).

\bibitem[{\citenamefont{Raphael and Lallement}(1968)}]{Raphael68}
\bibinfo{author}{\bibfnamefont{G.}~\bibnamefont{Raphael}} \bibnamefont{and}
  \bibinfo{author}{\bibfnamefont{R.}~\bibnamefont{Lallement}},
  \bibinfo{journal}{Solid State Commun.} \textbf{\bibinfo{volume}{6}},
  \bibinfo{pages}{383} (\bibinfo{year}{1968}).

\bibitem[{\citenamefont{Le et~al.}(2010)\citenamefont{Le, McEwen, Colineau,
  Griveau, and Eloirdi}}]{Le2010}
\bibinfo{author}{\bibfnamefont{M.~D.} \bibnamefont{Le}},
  \bibinfo{author}{\bibfnamefont{K.~A.} \bibnamefont{McEwen}},
  \bibinfo{author}{\bibfnamefont{E.}~\bibnamefont{Colineau}},
  \bibinfo{author}{\bibfnamefont{J.-C.} \bibnamefont{Griveau}},
  \bibnamefont{and} \bibinfo{author}{\bibfnamefont{R.}~\bibnamefont{Eloirdi}},
  \bibinfo{journal}{Phys. Rev. B} \textbf{\bibinfo{volume}{82}},
  \bibinfo{pages}{155136} (\bibinfo{year}{2010}).

\bibitem[{\citenamefont{Klimczuk et~al.}(2004)\citenamefont{Klimczuk, Gupta,
  Lawes, Ramirez, and Cava}}]{Klimczuk04}
\bibinfo{author}{\bibfnamefont{T.}~\bibnamefont{Klimczuk}},
  \bibinfo{author}{\bibfnamefont{V.}~\bibnamefont{Gupta}},
  \bibinfo{author}{\bibfnamefont{G.}~\bibnamefont{Lawes}},
  \bibinfo{author}{\bibfnamefont{A.~P.} \bibnamefont{Ramirez}},
  \bibnamefont{and} \bibinfo{author}{\bibfnamefont{R.~J.} \bibnamefont{Cava}},
  \bibinfo{journal}{Phys. Rev. B} \textbf{\bibinfo{volume}{70}},
  \bibinfo{pages}{094511} (\bibinfo{year}{2004}).

\bibitem[{\citenamefont{Shick et~al.}(1999)\citenamefont{Shick, Liechtenstein,
  and Pickett}}]{shick99}
\bibinfo{author}{\bibfnamefont{A.~B.} \bibnamefont{Shick}},
  \bibinfo{author}{\bibfnamefont{A.~I.} \bibnamefont{Liechtenstein}},
  \bibnamefont{and} \bibinfo{author}{\bibfnamefont{W.~E.}
  \bibnamefont{Pickett}}, \bibinfo{journal}{Phys. Rev. B}
  \textbf{\bibinfo{volume}{60}}, \bibinfo{pages}{10763} (\bibinfo{year}{1999}).

\bibitem[{\citenamefont{Shick and Pickett}(2001)}]{shick01}
\bibinfo{author}{\bibfnamefont{A.~B.} \bibnamefont{Shick}} \bibnamefont{and}
  \bibinfo{author}{\bibfnamefont{W.~E.} \bibnamefont{Pickett}},
  \bibinfo{journal}{Phys. Rev. Lett.} \textbf{\bibinfo{volume}{86}},
  \bibinfo{pages}{300} (\bibinfo{year}{2001}).

\bibitem[{\citenamefont{Moore and van~der Laan}(2009)}]{KMoore2009}
\bibinfo{author}{\bibfnamefont{K.}~\bibnamefont{Moore}} \bibnamefont{and}
  \bibinfo{author}{\bibfnamefont{G.}~\bibnamefont{van~der Laan}},
  \bibinfo{journal}{Rev. Mod. Phys.} \textbf{\bibinfo{volume}{81}},
  \bibinfo{pages}{235} (\bibinfo{year}{2009}).

\bibitem[{\citenamefont{Kolorenc et~al.}(2012)\citenamefont{Kolorenc,
  Poteryaev, and Lichtenstein}}]{kolorenc2012}
\bibinfo{author}{\bibfnamefont{J.}~\bibnamefont{Kolorenc}},
  \bibinfo{author}{\bibfnamefont{A.}~\bibnamefont{Poteryaev}},
  \bibnamefont{and} \bibinfo{author}{\bibfnamefont{A.~I.}
  \bibnamefont{Lichtenstein}}, \bibinfo{journal}{Phys. Rev. B}
  \textbf{\bibinfo{volume}{85}}, \bibinfo{pages}{235136}
  (\bibinfo{year}{2012}).

\bibitem[{\citenamefont{Shick et~al.}(2009)\citenamefont{Shick, Kolorenc,
  Lichtenstein, and Havela}}]{shick09}
\bibinfo{author}{\bibfnamefont{A.}~\bibnamefont{Shick}},
  \bibinfo{author}{\bibfnamefont{J.}~\bibnamefont{Kolorenc}},
  \bibinfo{author}{\bibfnamefont{A.~I.} \bibnamefont{Lichtenstein}},
  \bibnamefont{and} \bibinfo{author}{\bibfnamefont{L.}~\bibnamefont{Havela}},
  \bibinfo{journal}{Phys. Rev. B} \textbf{\bibinfo{volume}{80}},
  \bibinfo{pages}{085106} (\bibinfo{year}{2009}).

\bibitem[{\citenamefont{Daghero et~al.}(2012)\citenamefont{Daghero, Tortello,
  Ummarino, Griveau, Colineau, Eloirdi, Shick, Kolorenc, Lichtenstein, and
  Caciuffo}}]{daghero12}
\bibinfo{author}{\bibfnamefont{D.}~\bibnamefont{Daghero}},
  \bibinfo{author}{\bibfnamefont{M.}~\bibnamefont{Tortello}},
  \bibinfo{author}{\bibfnamefont{G.~A.} \bibnamefont{Ummarino}},
  \bibinfo{author}{\bibfnamefont{J.-C.} \bibnamefont{Griveau}},
  \bibinfo{author}{\bibfnamefont{E.}~\bibnamefont{Colineau}},
  \bibinfo{author}{\bibfnamefont{R.}~\bibnamefont{Eloirdi}},
  \bibinfo{author}{\bibfnamefont{A.~B.} \bibnamefont{Shick}},
  \bibinfo{author}{\bibfnamefont{J.}~\bibnamefont{Kolorenc}},
  \bibinfo{author}{\bibfnamefont{A.~I.} \bibnamefont{Lichtenstein}},
  \bibnamefont{and} \bibinfo{author}{\bibfnamefont{R.}~\bibnamefont{Caciuffo}},
  \bibinfo{journal}{Nat. Comm.} \textbf{\bibinfo{volume}{3}},
  \bibinfo{pages}{786} (\bibinfo{year}{2012}).

\end{thebibliography}

\end{document}